# An unexpectedly rapid decline in the X-ray afterglow emission of long gamma-ray bursts

G. Tagliaferri[1], M. Goad[2], G. Chincarini[1,3], A. Moretti[1], S. Campana[1], D.N. Burrows[4], M. Perri[5], S.D. Barthelmy[6], N. Gehrels[6], H. Krimm[6,7], T. Sakamoto[6,8], P. Kumar[9], P.I. Mészáros[4], S. Kobayashi[4], B. Zhang[10], L. Angelini[6,11], P. Banat[1], A.P. Beardmore[2], M. Capalbi[5], S. Covino[1], G. Cusumano[12], P. Giommi[5], O. Godet[2], J.E. Hill[4], J.A. Kennea[4], V. Mangano[12], D.C. Morris[4], J.A. Nousek[4], P.T. O'Brien[2], J.P. Osborne[2], C. Pagani[1,4], K.L. Page[2], P. Romano[1], L. Stella[13], A. Wells[2]

[1] INAF – Osservatorio Astronomico di Brera, Via Bianchi 46, I-23807 Merate, Italy
[2] Department of Physics and Astronomy, University of Leicester, Leicester, UK
[3] Università degli studi di Milano-Bicocca, P.za delle Scienze 3, I-20126 Milano,
[4] Department of Astronomy & Astrophysics, Pennsylvania State University, PA 16802, USA
[5] ASI Science Data Center, Via Galileo Galilei, I-00044 Frascati, Italy
[6] NASA Goddard Space Flight Center, Greenbelt, MD 20771, USA
[7] Universities Space Research Association, 10227 Wincopin Circle, Columbia, MD 21044, USA
[8] National Research Council, 2101 Constitution Avenue, NW, Washington, DC 20418, USA
[9] Department of Astronomy, University of Texas, RLM 15.308, Austin, TX 78712-1083, USA
[10] Department of Physics, University of Nevada, BOX 454002, Las Vegas, NV 891, USA
[11] Department of Physics and Astronomy, Johns Hopkins University, Baltimore, MD21218, USA
[12] INAF-Istituto di Astrofisica Spaziale e Cosmica, Via Ugo La Malfa 153, 90146 Palermo, Italy
[13] INAF-Osservatorio Astronomico di Roma, Via di Frascati 33, I-00040 Monteporzio, Italy

*NOTE: this paper has been accepted for publication in Nature, but it is embargoed for discussion in the popular press until formal publication in Nature.*

**Long gamma-ray bursts (GRBs) are commonly accepted to originate in the explosion of particularly massive stars, which gives rise to a highly relativistic jet. Internal inhomogeneities in the expanding flow give rise to internal shock waves that are believed to produce the gamma-rays we see[1,2]. As the jet travels further outward into the surrounding circumstellar medium 'external' shocks give rise to the afterglow emission seen in the X-ray, optical and radio bands[1,2]. Here we report on the early phases of the X-ray emission of five GRBs. Their X-ray light curves are characterised by a rapid fall-off for the first few hundred seconds, followed by a less rapid decline lasting several hours. This steep decline, together with detailed spectral properties of two particular bursts, shows that violent shock interactions take place in the early jet outflows.**

The GRB prompt γ-ray emission usually goes through a strong spectral evolution with the peak of the emission rising to higher energies in the early phases and then moving to lower energies[3], while the subsequent afterglow phase has an X-ray spectrum that is well represented by a power law model with an energy index of ~1 (e.g. [4]). The transition from the prompt γ-ray to the afterglow emission is expected to occur in the first few minutes following a GRB (see[5] and references therein). The multiwavelength observations of this transition and the early afterglow emission provide very important information regarding the properties and composition of the material released in these explosions, thus providing insight into the nature of the central engine[6]. Until now, however, this crucial time interval was largely unexplored (a few GRB afterglows at early times were observed, but with limited statistics[7,8,9,10], while for a few other bursts a very early optical emission was detected, e.g.[11,12]). With the successful launch of *Swift*[13] in November 2004, the situation has dramatically improved. We are now able to study this early afterglow phase starting a few tens of seconds after the burst explosion[14,15].

For five of the first seven GRBs promptly repointed by *Swift* the X-ray light curve, as seen by the X-Ray Telescope[16] (XRT) on board *Swift*, faded very fast (see Fig. 1 and its caption). GRB050126 and GRB050219a are the first two bursts with an X-ray light curve well sampled by XRT, allowing us a detailed investigation of their properties. They were detected and located by the Burst Alert Telescope[17] (BAT) on board *Swift* on January 26 and February 19, respectively[18,19]. In both cases *Swift* promptly slewed to the BAT burst locations and the XRT immediately began taking data, detecting bright and rapidly fading X-ray counterparts. In Fig. 2 we plot the BAT 20-150 keV light curve of these two bursts and the very early phases of the associated X-ray sources seen by XRT in the 0.2-10 keV band. The most striking feature of these X-ray light curves is their very steep initial decline, followed by a flattening a few hundred seconds later which is well represented by a broken power law model (see Table 1). The light curve of GRB050219a is well sampled and besides the general decay it clearly shows rapid variability on a time scale of a few tens of seconds, that in any case does not affect the general trend. In the following, for simplicity, we will refer to these X-ray sources as the afterglows, though we note that the early X-ray emission may instead be associated with the prompt emission from the burst. We will return to this subject later.

We sought for a possible delay of the afterglow onset by fitting the two X-ray light curves with a single power law model $\propto (t-t_0)^{-\alpha}$ (where $t_0$ would be the onset of the afterglow). In both cases, the decaying light curves can be fitted if the onset of the afterglow is shifted to $t_0$ ~100 s after the burst trigger with a power law slope of ~1, as typical of previously observed afterglows. For both GRBs (for $t_0$ > 80 s) the decay index in the first few hundred seconds would be $\alpha \leq 1.5$ and the emission is consistent with synchrotron radiation in the forward shock. In this case the spectral index β and the temporal index α ($f_\nu \propto \nu^\beta t^{-\alpha}$) must obey the relation $\beta = p/2$ and $\alpha = (3\beta -1)/2$ (ref. 6, assuming that the cooling frequency is below 0.2 keV), which is indeed satisfied by these bursts ($p=2.7\pm0.6$ for GRB050126 and $p=2.2\pm0.3$ for GRB050219a). However, while in the case of GRB050126 the light curve does not allow us to clearly state if a $(t-t_0)$ is better than a broken power law model, for GRB050219a a broken power law provide definitively a better fit. Moreover, for this burst we detected the X-ray afterglow emission at least as early as 87 s from the trigger (see Fig. 2 and relative caption). Thus, although the maximum of the afterglow emission could be at 105 s, the afterglow onset is clearly occurring before. While an onset of the afterglow some time after the trigger can be expected, for GRB050219a our data are not consistent with a single power law for the early afterglow decay whenever it starts *($t_0$)*.

To investigate whether the prompt and the early afterglow emissions are related, we converted both the BAT and XRT count rates to flux in the 0.2-10 keV band using the conversion factors derived from the BAT and XRT spectral analyses. As shown in Fig. 2, for GRB050219a the BAT and XRT light-curves are discontinuous, whereas for GRB050126, although unlikely, this could still be possible. For GRB050219a the afterglow emission peaks about one hundred seconds after the burst trigger and this occurs after the prompt emission has already faded away. The spectral index during the burst for these GRBs is quite different from the spectrum during the early X-ray emission. However, given that burst spectra can go through strong spectral evolution, the XRT spectrum at the beginning could still be due, at least partly, to the prompt emission. Thus, we search for a possible spectral evolution of the X-ray sources detected by XRT. There is no significant evidence in either source, for a X-ray spectral change across the break (with some caution for GRB050126, see Table 1 and relative caption). Thus, given that the BAT and XRT spectra are very different, we have a clear indication, at least for GRB050219a, that the XRT source is due only to the afterglow emission. This property, together with the discontinuity in the lightcurve, suggests that the burst and the early afterglow emission are produced by different mechanisms. This conclusion is in agreement with the expectation that the prompt γ-ray radiation is produced in internal shocks whereas the afterglow radiation is produced in the external shock[6,23].

The prompt γ-ray and the early X-ray afterglow emission of GRB050219a and probably also of GRB050126 require at least two and possibly three distinct mechanisms as discussed below. For a self-similar forward shock solution in the standard GRB model the time shift between the γ-ray trigger and the onset of afterglow emission is expected to be small. If this is within a few tens of seconds of the burst trigger time, which our data seem to indicate at least for GRB050219a, then we have a very steep initial decline of the early X-ray afterglow lightcurve, which requires explanation. A rapidly falling X-ray light curve at early times may arise in a hot cocoon accompanying a relativistic jet[24,25] or could be the photospheric emission associated with the outflow from the explosion[26]. However, in the simplest versions of these models the spectrum of the emergent radiation is thermal, which is inconsistent with the power law spectrum observed for the two bursts. Some modifications to these models involving a Comptonized power law tail of thermal radiation, for instance, might produce the observed behavior. An alternative possibility is that the steep afterglow decay is produced in the external shock from a jet consisting of narrow regions of angular size $\leq \Gamma^{-1}$, where $\Gamma$ is the jet Lorentz factor. As $\Gamma$ decreases, the opening angle from which radiation can be seen becomes larger, without encompassing a larger fraction of the jet. A steep decay in the light curve is thus produced[27]. Yet another possibility is light delay effects in off-axis emission ($\theta > \Gamma^{-1}$) from a relativistic jet arriving at the observer when emission from $\theta < \Gamma^{-1}$ has dropped to very small values due to the adiabatic cooling of the shock heated shell[28].

A very interesting possibility is that the steep, early, X-ray lightcurve is due to emission from the reverse shock heated ejecta[28,29]. The peak of the synchrotron emission in the reverse shock is in the infrared or optical. These photons, if scattered by relativistic electrons in the ejecta, emerge in the X-ray band. The X-ray lightcurve in this case will decline roughly as $t^{-2.6}$, which is consistent with observations. However, in order to avoid a very bright early optical radiation from these bursts, which was not seen, the ejecta may need to be highly enriched with $e^{\pm}$ pairs, with an ejecta Lorentz factor of at least a few hundred.

We note that none of these models is completely consistent with all the available data in the γ-rays, X-rays, and optical upper limits for these two GRBs, which suggests the need for refining the current models.

______________________________________________________________________

______________________________________________________________________

Acknowledgements: The authors acknowledge support from ASI, NASA and PPARC.

Competing interests statement. The authors declare that they have no competing financial interest

Correspondence and requests for materials should be addressed to G.T. (tagliaferri@merate.mi.astro.it)

## Table 1: BAT time averaged spectral parameters

|  | GRB050126 | GRB050219a |
| --- | --- | --- |
| $T_{90}$(s) | 25.7 | 23.6 |
| Flunce[1] | $(1.7\pm0.3)\times10^{-6}$ | $(5.2\pm0.4)\times10^{-6}$ |
| Model | Power law | Cut-off power law |
| $\beta'$ | 0.34±0.14 | -0.75±0.30 |
| $E_{peak}$ (keV) |  | 90±9 |
| $\chi^2_{red}$/d.o.f. | 1.25/57 | 0.86/56 |

### XRT first orbit spectral fit and parameters

|  |  |  |
| --- | --- | --- |
| $N_{H\,Gal}$ (cm$^{-2}$) | $5.3\times10^{20}$ | $8.5\times10^{20}$ |
| $N_{H\,excess}$ (cm$^{-2}$) | - | $(2.25\pm0.60)\times10^{21}$ |
| $\beta$ | 1.26±0.22 | 1.1±0.2 |
| $\chi^2_{red}$/d.o.f. | 1.06/8 | 1.02/53 |

### XRT light curve fits

|  | GRB050126 | GRB050219a |
| --- | --- | --- |
| $\alpha_1$ | $2.52^{+0.50}_{-0.22}$ | 3.15±0.22 |
| Break (s) | $425^{+560}_{-120}$ | 307±26 |
| $\alpha_2$ | $1.00^{+0.17}_{-0.26}$ | 0.82±0.07 |
| C-stat/n.d.p. | 26.1/20 |  |
| $\chi^2_{red}$/d.o.f. |  | 1.41/39 |
| $t_0$ (s) | $105^{+9}_{-11}$ | 105±5 |
| $\alpha$ | 1.08±0.09 | 0.98±0.05 |
| C-stat/n.d.p. | 31.7/20 |  |
| $\chi^2_{red}$/d.o.f |  | 2.6/40 |

[1] erg cm$^{-2}$ in the band 15-350 keV.

**Table 1**. BAT and XRT best fit parameters. The BAT average spectrum of GRB050126 is well fitted by a simple power law ($f_E \propto E^{-\beta}$); while for GRB050219a a cut-off power law ($f_E \propto E^{-\beta}e^{(-E/E0)}$) is necessary. Moreover, for the latter we detect clear spectral evolution (see Table 2). For both GRBs, the XRT spectral analysis is performed on the spectra accumulated over the first orbit. We also checked for a spectral evolution across the break. In both cases a combined fit to the two spectra accumulated before and after the break with a single absorbed power law model provides a good fit with a $\chi^2_{red} \cong 1$. If we fit the two spectra separately, leaving also the $N_H$ free to vary for GRB050219a, we have an indication that the spectrum becomes harder after the break ($\alpha \sim 1.7$ for both GRBs), but the slopes are fully consistent at 90% confidence range. Moreover, if we tied

the $N_H$ to be the same in the fit of the two X-ray spectra associated with GRB050219a, the two spectral indexes differ by less than 0.1. Thus, for this GRB we can definitively say that there is no spectral evolution across the break. While, for GRB050126 the statistic after the break is such that we can not speculate much about its X-ray spectrum. For this source we would be able to see a spectral change only if $\Delta\alpha \geq 0.8$. The XRT light curves shown in Fig. 1, with the onset time coinciding with the onset of the prompt, can not be fitted by a simple power law model; an F-test shows that the chance probability for the improvement of the broken power law model is less than $10^{-4}$ and $10^{-8}$ for GRB050126 and GRB050219a, respectively. We also report the results of the best-fit for a single power law model where also the onset time ($t_0$) is fitted. All errors quoted in the Table are for a $\Delta\chi^2=2.71$. For the fit of the GRB050126 light curve we used the C-statistics, because of the small number of counts in each bin.

**Table 2 GRB050219a BAT spectral fits**

|  | Int. 1 (8s) | Int. 2 (6s) | Int. 3 (4s) | Int. 4 (7s) | Int 4+5 (14s) |
|---|---|---|---|---|---|
| β | $-2.0^{+0.7}_{-1.0}$ | $-1.35^{+0.55}_{-0.65}$ | $-0.70^{+0.35}_{-0.40}$ | $-0.65^{+0.40}_{-0.45}$ | $-0.35^{+0.40}_{-0.50}$ |
| $E_{peak}$(keV) | $84^{+13}_{-9}$ | $87^{+15}_{-10}$ | $127^{+50}_{-22}$ | $73^{+12}_{-7}$ | $68^{+13}_{-7}$ |
| $\chi^2_{red}$/d.o.f. | 1.01/56 | 0.77/56 | 0.92/56 | 0.69/56 | 0.94/56 |

**Table 2**. BAT best-fit parameters for the prompt spectra of GRB050219a. For this burst we find clear evidence of spectral evolution in the prompt emission light curve and perform the spectral analysis over 5 consecutive time intervals (see Fig. 2). Spectra for the first four intervals are well fitted by a cut-off power law that shows a hardening up to interval three and then a softening. The fifth interval does not have enough counts to constrain the parameters. Here we report the time duration and the spectral fit parameters for the first four intervals and for the fourth and fifth intervals added together.

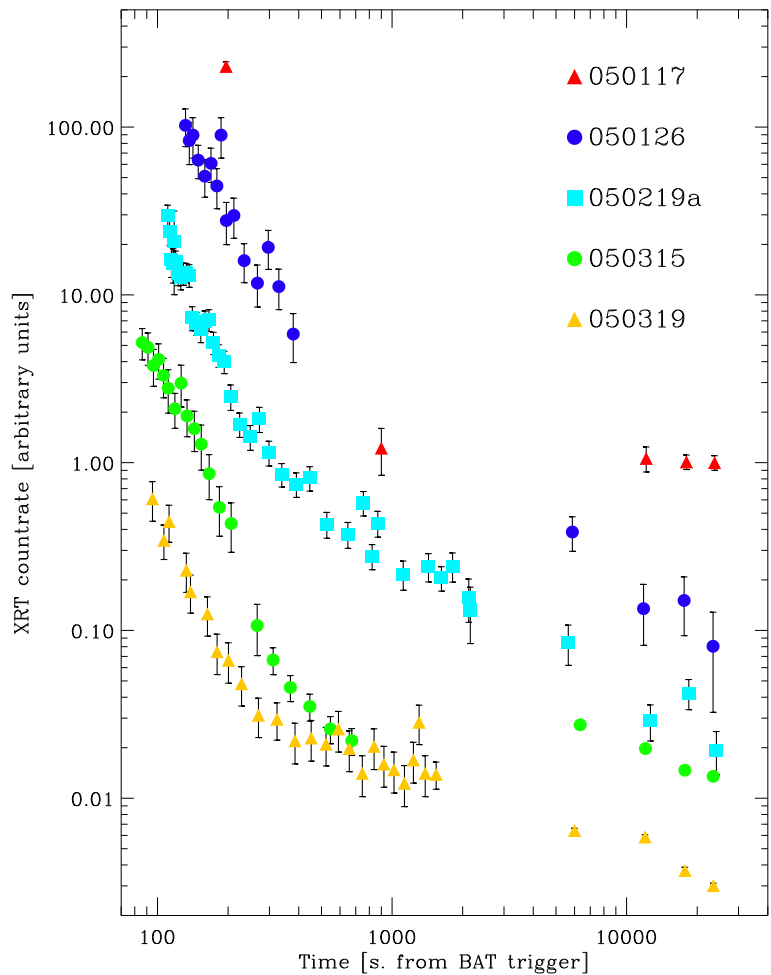

**Figure 1** The steep early X-ray light curves of five GRBs observed by XRT[16]. During the Swift[13] performance verification phase, that ended April 5th, seven GRBs, discovered by the BAT[17] on board Swift, were promptly repointed by the satellite. The XRT began taking data starting up to a few tens of seconds after the burst explosion. A bright and fading X-ray counterpart was always detected. For five of them, whose light curves are shown in this figure, the fading was very fast, flattening after a few hundred seconds. For each GRB the XRT count rates are rescaled by an arbitrary constant factor for clarity, while the error bars represent the standard deviation. GRB050126 and GRB050219a are the first two bursts with X-ray light curves well sampled by XRT. The UV-optical Telescope[20] (UVOT) on board Swift could not observe the field of GRB050126 due to the proximity of the bright star Vega. Four and a half hours after the burst a new IR source was detected in the Ks band by the Keck telescope within the XRT position error circle, with a subsequent redshift determination z=1.29 of the host galaxy[21]. For GRB050219a the UVOT did not find an optical counterpart to the X-ray source down to a limiting magnitude of V=20.7[22]. No optical/NIR or radio counterpart to this GRB has been reported.

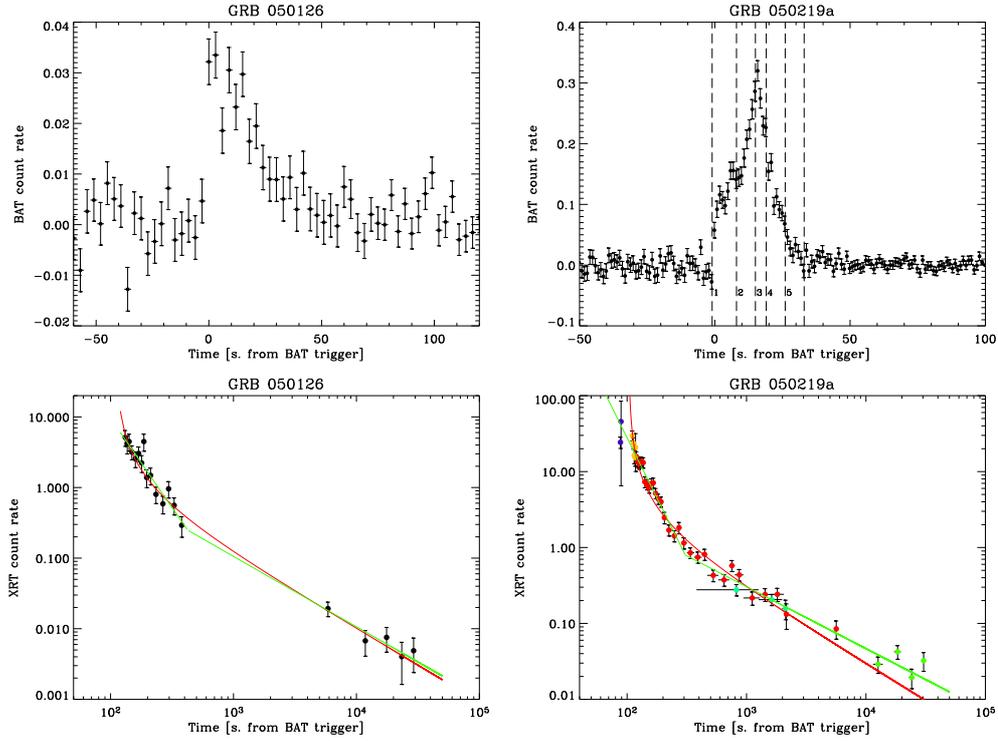

**Figure 2** The X-ray light curves of GRB050126 and GRB050219a as seen by BAT and XRT (error bars represent the standard deviation). On January 26, 2005, 12:00:54 UT, GRB050126 was detected and located by the BAT (top left panel). Swift promptly slewed to the burst and settled at 129 s after the BAT trigger. Then the XRT took data until 12:07:42, detecting a very bright and rapidly fading X-ray counterpart. GRB050126 was further observed by the XRT for the following 8 orbits (bottom left panel). On February 19, 2005, 12:40:01 UT, GRB050219a triggered the BAT (top right panel). Swift autonomously slewed to the BAT burst location and was on target after 87 seconds. The XRT executed the standard sequence of observations for GRBs[30], again detecting a very bright and rapidly fading X-ray counterpart across 5 orbits (bottom right panel). In all plots times are from the onset of the prompt emission. For GRB050219a this is 6 second earlier than the time reported in [19]. GRB050126 is a fast-rise-exponential-decay GRB with a total duration of ~30 s. GRB050219a has a more complex and multi-peaked light curve with a total duration of ~32 s. For this burst we performed the BAT spectral analysis over 5 intervals as shown. The bottom panels show the XRT light curves. The green lines represent a broken power law best fit to the afterglow decay, while the red lines represent the best fit for a model $\propto (t-t_0)^{-\alpha}$. For GRB050219a we also have an earlier detection before the decaying part (see blue points), which seems to indicate that the peak

is in between these two points (not used in the fit) and the decaying part of the light curve, i.e. in the range 88-110 s. The blue and yellow points are in Photodiode mode, the red points are in Window Timing mode and the green points in Photon Counting mode (see[30] for a description of the XRT operating modes).

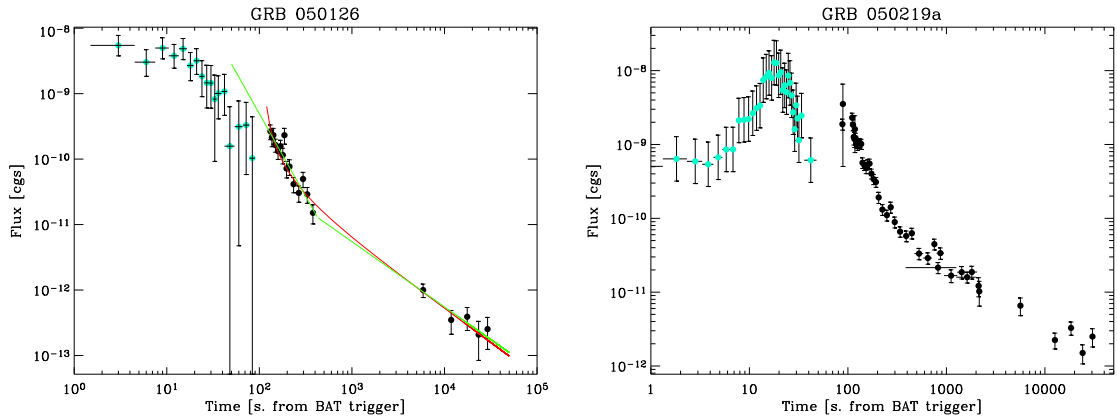

**Figure 3** Evolution of the two GRB X-ray light curves from the prompt to the afterglow phase. The BAT and XRT count rates are converted into fluxes in a common energy band (0.2-10 keV). The conversion factors have been calculated using the best fit models that reproduced the BAT prompt spectra (for GRB050219a we use the values reported in Table 2) and the XRT afterglow spectra, respectively. The error bars represent the standard deviation plus the estimated uncertainties in the conversion factors. For GRB050219a the X-ray source detected by XRT is at a higher level than the late stages of the prompt emission: there is a clear discontinuity between the BAT and XRT lightcurves. The XRT lightcurve also shows a hint of a rising phase before the onset of the decay. Note that the BAT detector is taking data all the time. We stop plotting them after ~80 and 50 s, for the two GRBs respectively, because the sources are not detected any more. For a 5σ detection of GRB050219a X-ray source seen by XRT at ~90-100 s, BAT would need more than 100 s. Given that this source is rapidly fading, it is to weak to be detected by BAT. For GRB050126 the X-ray flux is weaker and the discontinuity is not so evident. Moreover, the BAT conversion factor is calculated over the averaged spectrum. For a strong spectral evolution from hard to soft the latter BAT points would have higher fluxes. However, the 5 s peak spectrum and the averaged total spectrum have very similar spectral indexes[18], so we do not have indication of a strong spectral evolution. In conclusion for GRB050126 the BAT and XRT light curves do not seem to simply connect as well, although for this GRB this cannot be ruled out.